# THE IMPACT OF THE PHARMACEUTICAL INDUSTRY ON THE INNOVATION PERFORMANCE OF EUROPEAN COUNTRIES

**Szabolcs Nagy[1], Sergey U. Chernikov[2], Ekaterina Degtereva[3]**

**Faculty of Economics, University of Miskolc, Hungary**[1]

**Faculty of Economics, The People's Friendship University of Russia (RUDN University), Russia**[2,3]

**Summary:** *There are significant differences in innovation performance between countries. Additionally, the pharmaceutical sector is stronger in some countries than in others. This suggests that the development of the pharmaceutical industry can influence a country's innovation performance. Using the Global Innovation Index (GII) and selected performance measures of the pharmaceutical sector, this study examines how the pharmaceutical sector influences the innovation performance of countries from the European context. The dataset of 27 European countries was analysed using simple, and multiple linear regressions and Pearson's correlation. Our findings show that only three indicators of the pharmaceutical industry– pharmaceutical Research and Development (R&D), pharmaceutical exports, and pharmaceutical employment–explain the innovation performance of a country largely. Pharmaceutical R&D and exports have a significant positive impact on a country's innovation performance, whereas employment in the pharmaceutical industry has a slightly negative impact. Additionally, global innovation performance has been found to positively influence life expectancy. We further outline the implications and possible policy directions based on these findings.*

*Acknowledgement: The article was prepared with the financial support of the RFFR as part of a research project 'Opportunities and prospects for the development of strategic alliances of innovative organizations in Hungary and Russia in the field of biotechnology and pharmaceuticals', project № 21-510-23004.*

## 1. Introduction

European citizens can expect to live 30 years longer than they did a century ago and have a better quality of life, thanks to advances in science and technology and the research-based pharmaceutical industry (EFPIA 2021). However, some diseases, such as various cancers, rare diseases, and Alzheimer's still cannot be treated or are inadequately treated. Further, the recent emergence of COVID -19 and its mutations has become a major threat today.

Innovation is one of the main drivers of GDP growth and a key success factor in global competition. The pharmaceutical industry is one such industry that is highly innovative. According to the 2020 EU Industrial R&D Investment Scoreboard, the pharmaceutical industry has the highest ratio of R&D investment to net sales (15.4%), followed by 'software and computer services' (11.8%) and technology hardware and equipment (9.00%) (Grassano et al 2020).

However, previous studies have not addressed how the pharmaceutical industry affects the innovation performance of countries; thus, it is timely and crucial to examine the relationship between the key performance indicators of the pharmaceutical industry and country-level innovation performance. We hypothesise that a stronger pharmaceutical industry positively influences a country's innovation performance.

The main objective of this study is to investigate how the pharmaceutical industry affects country-level innovation performance in the European context. To achieve this objective, the relationship between the Global Innovation Index and pharmaceutical industry metrics were examined. This study also analyses the relationship between innovation performance and life expectancy to add to the overall body of academic literature.

## 2. Literature review
### 2.1. Innovation and the Global Innovation Index (GII)

Innovation plays a significant role, contributing to a country's economic growth (Maradana et al, 2017; Pece et al, 2015). The need to assess innovation performance and comparability across countries has led to the development of the Global Innovation Index (Cornell University, INSEAD, and WIPO 2020), recognised by the European Union. The Global Innovation Index provides a framework for measuring innovation that enables policymakers to develop their strategies.

The tendency to increase the volume of innovation financing, as one of the strongest engines of the modern economy, is not new (Kuznets 1972). Cornell University, in cooperation with the French business school INSEAD and the World Intellectual Property Organization, developed the GII to determine the relationship and assess the innovativeness of economies. Since innovation is a subjective indicator, the GII approach is promising and relatively objective (Iqbal and Rahman 2020). The annual report assesses the innovation of the economies of all countries based on a number of indicators for compiling the rating. Proxies that may not reflect innovation in their true sense are used because of the lack of data on innovation in some countries. The report also examines leading scientific and technical clusters, which are centres of innovation activity in countries. Each country has its own weaknesses and strengths, determined by the analysis of institutions, human resources and their development, infrastructure, the degree of development of the market and business, and the technological and creative results of innovation activity.

The purpose of this report is to provide meaningful data on innovations that can help countries assess the effectiveness of their innovation activities and make decisions on further development plans (Androschuk 2021). It is possible to combine data from previous years and conduct a trend analysis since the report is published annually (Naqvi 2016). Overall, the index reflects the innovation hierarchy of the world quite well.

The GII ranks economies according to their capacity for innovation. The multidimensional index comprises approximately 80 indicators, divided into innovation inputs and outputs. The overall GII score is based on the average of the innovation input and output sub-indices. Both sub-indices have equal weights for the overall GII score. The innovation input sub-index, which captures elements of the economy that enable innovative activities, comprises the following five pillars: institutions, human capital and research, infrastructure, market sophistication, and business sophistication. These elements foster innovation activities in a country. The innovation output sub-index, which shows the outcome of innovative activities within the economy, has two pillars: knowledge, technology, and creative output. Each pillar was divided into three sub-pillars (Figure 1).

In 2020, 131 countries were studied, representing 93.5% of the total population and 97.4% of global GDP. The GII has become both a primary reference for innovation and an 'action tool' for countries to incorporate the GII into their innovation agenda.

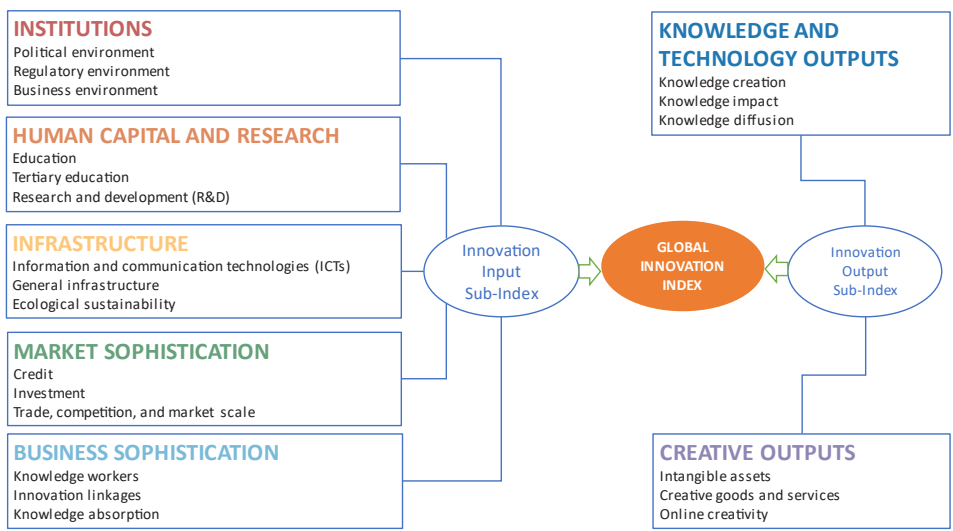

**Figure 1** – Framework of the Global Innovation Index. Source: based on Cornell University, INSEAD, and WIPO (2020, p. 205)

The GII 2020 identified six key findings (Cornell University, INSEAD, and WIPO, 2020). The first key finding is that the pandemic COVID-19 has had a positive effect on innovation. Innovation has increased in the pharmaceutical and biotechnology sectors owing to the development of vaccines and advances in education and retail. COVID-19 has also had a positive impact on the pharmaceutical industry in general (Daniel et al 2021). Governments have sought to mitigate the negative impacts of innovation in sectors affected by the pandemic. Second, the crisis led to a decline in funding for innovation. Venture capital deals have also declined, and initial public offerings (IPOs) and start-ups have grown less. It is predicted that venture capital will take longer to recover than R&D; thus, there may be a shortage of funding for innovation. Innovation diffusion is directed toward health, online education, big data, e-commerce, and robotics. Third, the global innovation landscape is shifting eastward, with China, Vietnam, India, and the Philippines showing an increasing trend in the GII growth. China, as part of the middle-income group of countries, is an innovation leader and stands out for producing innovations that match those of the top ten economies.

Another important insight is that developing countries can excel in certain pillars of innovation. It is worth observing the world's top rankings in certain aspects of innovation in the GII, such as venture capital, R&D, entrepreneurship or high-tech manufacturing. The balance of the innovation system was also assessed in countries with the GII. The GII is also used to determine the innovation performance in relation to the level of economic development. The fifth important finding is the existence of regional differences, with some countries having a great innovation potential. North America and Europe are at the top, followed by Southeast Asia, East Asia, and Oceania. Finally, innovation in some high-income economies, notably China, is concentrated in science and technology clusters. The US has the most clusters (25), followed by China (17), Germany (10), and Japan (5) (Cornell University, INSEAD, and WIPO 2020).

It is also important to note that most economies that have moved up the GII over time have benefited greatly from their integration into value chains and innovation networks. Today, international openness and collaboration in innovation pose real risk. However, the joint search for medical solutions during the pandemic showed how powerful collaboration could be. The speed and efficiency of this collaboration show that internationally coordinated R&D missions can effectively counter the tendency toward increased isolationism and address important societal issues.

According to the GII 2020, Switzerland is the most innovative economy in the world, with an index score of 66.1 (Appendix 1). Switzerland's success is the result of its production of high-value innovations, knowledge-intensive employment, and high R&D spending. Switzerland was followed by Sweden (62.5) and the United States (60.6). The top ten also include the United Kingdom (59.8), the Netherlands (58.8), Denmark (57.5), Finland (57), Singapore (56.6), Germany (56.5), and the Republic of Korea (56.1).

Countries at the bottom of the GII ranking are Zambia (19.4), Mali (19.2), Mozambique (18.7), Togo (18.5), Benin (18.1), Ethiopia (18.1), Niger (17.8), Myanmar (17.7), Guinea (17.3), and Yemen (13.6). In Africa, low scores in the GII can be explained by limited access to innovation systems due to the low level of scientific and technological activities, dependence on government or foreign funders as a source of R&D, low business absorptive capacity, and a difficult business environment (Cornell University, INSEAD, and WIPO 2020).

### 2.2. Innovation and the pharmaceutical industry

The pharmaceutical industry has long relied more on research and innovation than other industries (Cardinal 2001, Romasanta et al 2020). R&D is crucial for the growth and future success of pharmaceutical research companies (Schuhmacher and Kuss 2018). Valuable pharmaceutical innovations should be encouraged, identified, and rewarded. A medicine can only be considered "truly innovative" if it offers additional clinical efficacy and/or effectiveness compared to the current treatment. In addition, it is 'valuable' only if it addresses an unmet medical need (Annemans et al 2011). Investments in pharmaceuticals are of a long-term nature, making it difficult in assessing the effectiveness of investments since the results of research and development can only be obtained in the distant future. At the same time, methodological disputes continue on accounting for numerous hidden costs that can further increase the cost of research. Even further, the development of new drugs is becoming increasingly expensive; estimates range from USD 1 billion to USD 11 billion (Kumar and Sundarraj 2018). Consequently, the key strategies used in the field of scientific research and innovation in the pharmaceutical industry today are mergers and acquisitions, licensing, and strategic partnerships (Haraszkiewicz-Birkemeier and Hołda 2019). The innovation and investment activities of pharmaceutical companies have a significant impact on the development of society as well as globalisation. With existing and new diseases to fight, if a pharmaceutical company wants to stand out, it has to spend more on R&D (Yildirim and Mestanoğlu 2019). According to McKinsey experts' estimates, the cost of innovation accounts for up to half of pharmaceutical companies' profits, well above for the same indicator in other industries. However, high R&D spending is not restricted to the pharmaceutical industry (Brennan et al. 2020). This is also high in the high-tech, media, and telecom industries, as well as in the automotive and assembly industries (Figure 2).

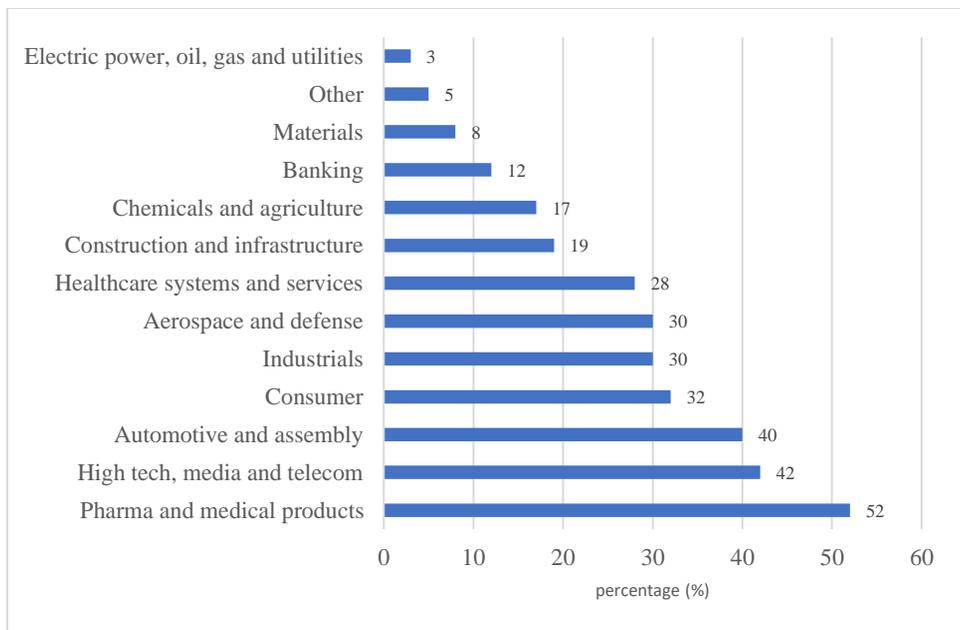

**Figure 2** Global private-sector R&D spending as a share of EBITDA by industry percentage, Source: Brennan et al. (2020)

At the same time, the pharmaceutical industry is striving to reduce the ever-increasing costs of research and development using artificial intelligence, which provides a better understanding of the relationships between various formulas and parameters (Krishnaveni et al. 2020). However, with rapidly growing development costs, the reduced profitability of new medical organisations and missed breakthrough innovations can negatively affect the future of the pharmaceutical industry (van

Vierssen Trip, Nguyen, and Bosch 2017). Therefore, one of the tasks of pharmaceutical companies is to find a compromise between innovation and pricing (Konopielko and Trechubova 2019).

Medical progress is driven by the pharmaceutical industry and vice versa. The aim is to provide patients with innovative treatments that are widely available and accessible by turning fundamental research into new products in the market (EFPIA 2021).

### 2.3. Pharmaceutical industry in Europe

The pharmaceutical industry is one of the most powerful high-tech industries in Europe and is a key asset for the European economy. It is the main driver of current and future growth as well as global competitiveness. The pharmaceutical industry in Europe employs approximately 830,000 people directly and approximately 2.5 million people indirectly (EFPIA 2021). The R&D spend for Europe alone was 39,000 million EUR, a huge investment. However, this industry also has several problems. However, additional regulatory hurdles, rising R&D costs, and the impact of fiscal austerity measures, since 2010 are the biggest challenges currently facing the European pharmaceutical industry (EFPIA 2021). For example, the estimated R&D cost for a new drug in 2014 was 1,926 million EUR, 2.5 times higher than that in the 1990s, and 14.3 times higher than that in the 1970s. (DiMasi et al. 2016) Nowadays, the time to market is also very long, 12-13 years on average. In addition, rapidly developing countries, particularly Brazil, China, and India, pose a serious threat to the fragmented EU pharmaceutical market by siphoning off R&D investment and activity, including the long-standing, strong market dominance of the United States (EFPIA 2021).

Key indicators are presented based on the latest report of the European Federation of Pharmaceutical Industries and Associations (EFPIA 2021) to provide an overview of the European pharmaceutical industry, representing 36 national pharmaceutical industry associations, 39 leading pharmaceutical companies, and 14 small-and medium-sized enterprises (e.g. the research-based pharmaceutical industry operating in Europe). The indicators comprise pharmaceutical industry R&D spending, pharmaceutical production, pharmaceutical employment, pharmaceutical market value, pharmaceutical exports, and pharmaceutical imports. The figures are from 2019..

### 2.3.1.  Pharmaceutical R&D

In 2019, the pharmaceutical industry in Europe invested more than 37,700 million EUR into R&D. The distribution of investments is uneven with more than 22.2% of this amount being invested in Germany. Substantial R&D investments also characterise the pharmaceutical sector in Europe, Switzerland, the UK, France, and Belgium. For example, in Russia, this amount is much lower at only 727 million EUR. In Croatia, at the other end of the spectrum, it is only 40 million; that is, pharmaceutical R&D in Germany is almost 212 times higher than in Croatia (Figure 3).

However, globally far too little has been invested in sustainable research and development (R&D) to prevent public health crises as the outbreak of the COVID-19 pandemic showed (Eyal-Cohen and Rutschman 2020).

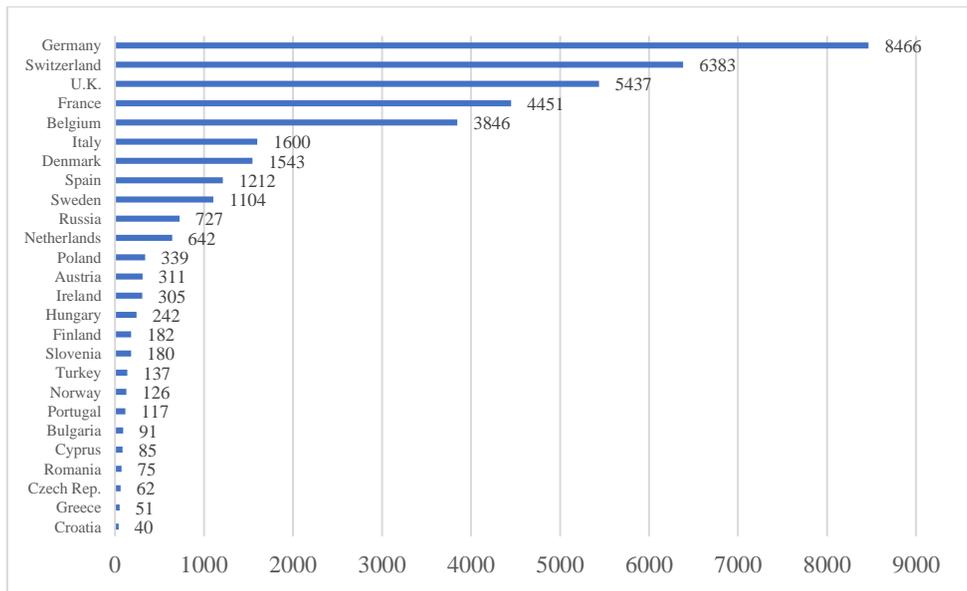

**Figure 3** Pharmaceutical industry research & development in Europe (in million EUR, 2019), Source: based on EFPIA (2021, p.7)

### 2.3.2. *Pharmaceutical production*

Pharmaceutical production is not uniform in Europe. Switzerland is the largest pharmaceutical producer, with an output of 54,305 million EUR, followed by France, Italy, Germany, and the UK. These countries are the top five pharmaceutical producers in Europe. By contrast, Iceland, Bulgaria, Cyprus, Latvia, and Slovakia were the smallest European producers (Figure 4).

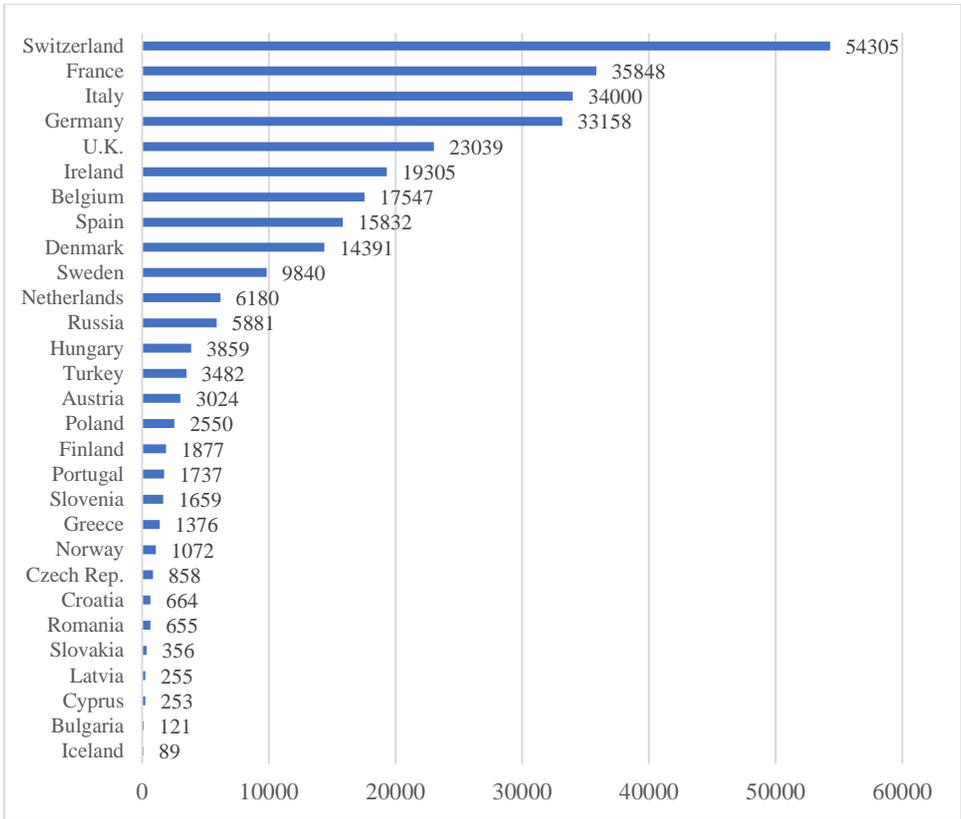

**Figure 4** Pharmaceutical production in Europe (in million EUR, 2019), Source: based on EFPIA (2021, p.11)

### 2.3.3. *Pharmaceutical employment*

Europe's pharmaceutical industry is a major industrial employer in the high-technology sector. It employs approximately 830,000 people, almost 15% of whom are in Germany alone. France is also a major employer, with almost 99,000 people employed in the pharmaceutical industry. However, Italy, Spain, and Switzerland also have many people working in this industry. Estonia has the smallest number of employees in Europe, followed by Iceland, Malta, Lithuania, and Cyprus (see Figure 5).

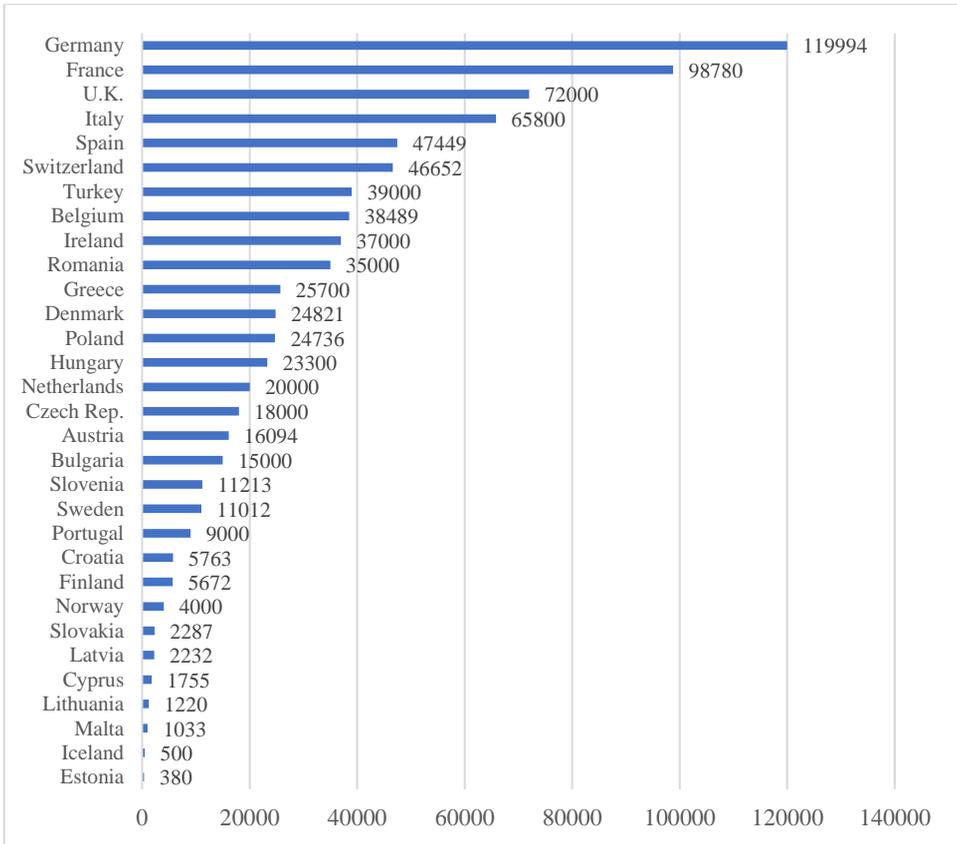

**Figure 5** Employment in the pharmaceutical industry (in units, 2019), Source: based on EFPIA (2021, p.12)

### 2.3.4. *Pharmaceutical market value*

Pharmaceutical market value refers to the sale of medicines for human use at ex-factory prices through all distribution channels (pharmacies, hospitals, dispensing doctors, supermarkets, etc.), whether dispensed on prescription or at the request of the patient. Market value does not include the sales of veterinary medicines. Germany is the largest market in Europe, with a share of 17.79% (Figure 6). This is followed by France (12.89%), Italy (10.60%), the UK (10.24%), Russia (7.77%), and Spain (7.52%). In contrast, the smallest market in Europe is Iceland, with less than 0.06% share, followed by Cyprus, Malta, Estonia, and Latvia, all with less than 0.20% share (Figure 6).

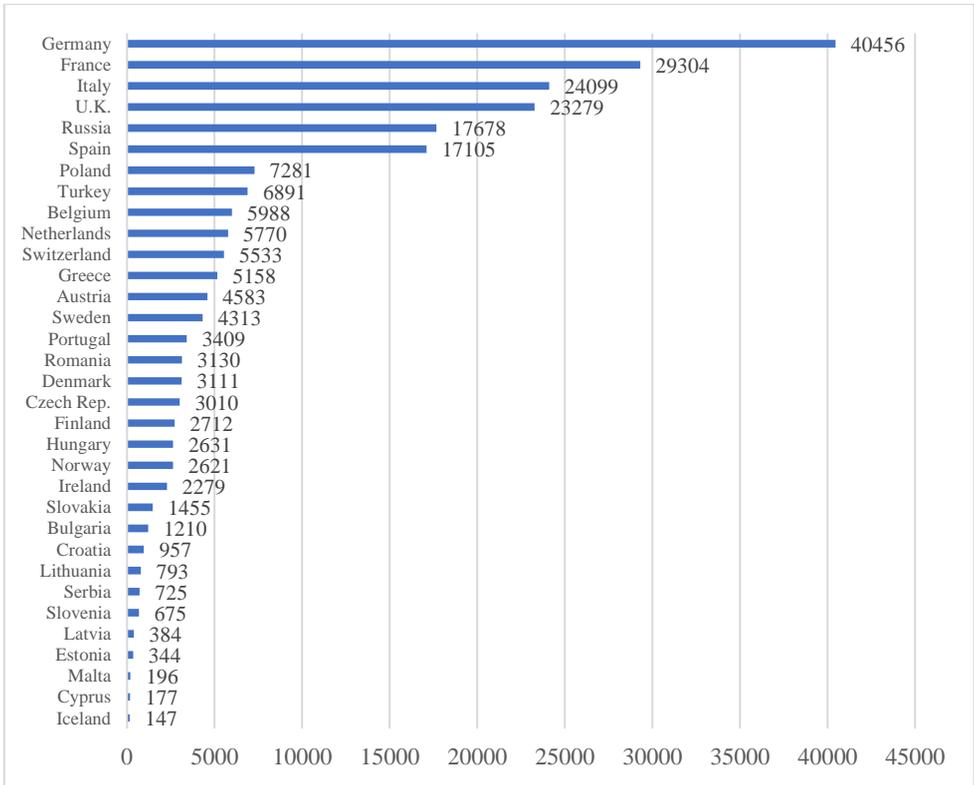

**Figure 6** Pharmaceutical market value (at ex-factory prices, in million EUR, 2019), Source: based on EFPIA (2021, p.15)

### 2.3.5. *Pharmaceutical exports*

Germany is the largest exporter of pharmaceuticals in Europe, closely followed by Switzerland (Figure 7). The top five exporting countries were Belgium, Ireland, and the Netherlands. Estonia has modest export performance, followed by Luxembourg, Malta, Cyprus, and Russia. These countries had the worst pharmaceutical export performances.

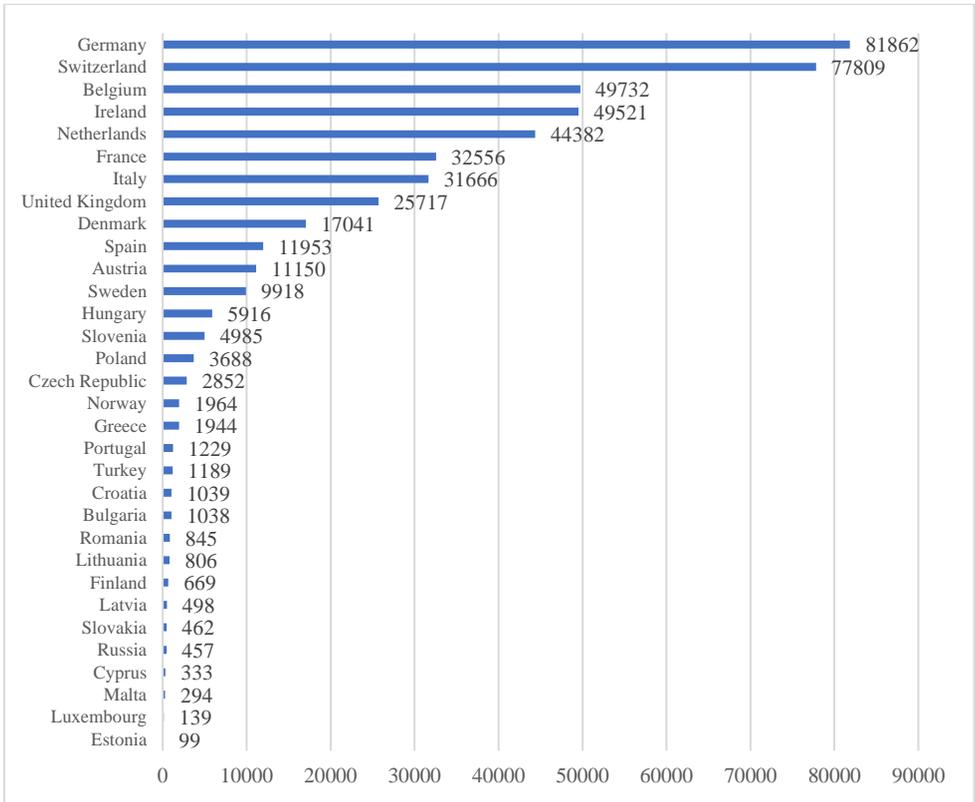

**Figure 7** Pharmaceutical exports (in million EUR, 2019), Source: based on EFPIA (2021, p.18)

### 2.3.6. *Pharmaceutical imports*

Germany was the largest importer of pharmaceuticals in Europe, followed by Belgium. Figure 8 shows that the Netherlands, Switzerland, and Italy were among the top five EU importers. Malta's pharmaceutical imports were the smallest in Europe, followed by Cyprus, Luxembourg, Estonia, and Latvia.

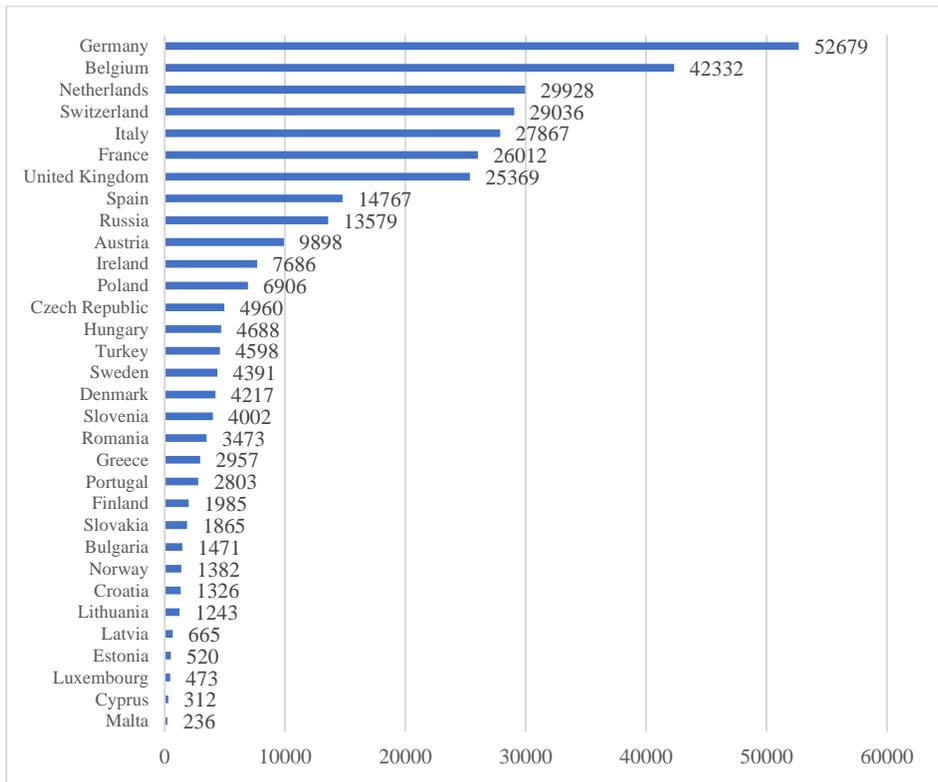

**Figure 8** Pharmaceutical imports (in million EUR, 2019), Source: based on EFPIA (2021, p.19)

## 3. Research methodology and sample

We analysed the relationship between the GII and the key performance indicators of the pharmaceutical industry to achieve the research objectives. The GII measures the innovation performance of 131 countries (Cornell College, INSEAD, and WIPO, 2020). The study of the European Federation of Pharmaceutical Industries and Associations contains pharmaceutical industry data of 32 countries (EFPIA 2021). It must be noted that the GII does not contain any sub-indices specific to the pharmaceutical industry.

Data were missing for some indicators (R&D, production, and employment) in EFPIA (2021). Therefore, countries with missing data (Estonia, Iceland, Latvia, Lithuania, Malta, and Slovakia) were excluded from the analysis. Employment data for Russia were not available in EFPIA (2021) and were taken from Statista (2017).

The raw dataset for the remaining 27 countries can be found in Appendix 2. The sample consisted of 27 European countries: Austria, Belgium, Bulgaria, Croatia, Cyprus, Czech Republic, Denmark, Finland, France, Germany, Greece, Hungary, Ireland, Italy, Netherlands, Norway, Poland, Portugal, Romania, Russia, Slovakia, Slovenia, Spain, Sweden, Switzerland, Turkey, and the United Kingdom. The relatively small sample size of 27 was one of the limitations of this study. Another limitation is that the results can only be interpreted in a European context, and the data refer to only one year (2019), since only the most recent EFPIA figures were used for the analysis.

We used multiple linear regression (MLR) to test the working hypothesis that a more developed pharmaceutical sector positively affects a country's innovation performance. In the simple linear regression, the global innovation index score (GII) was the dependent variable, while pharmaceutical research and development (R&D), pharmaceutical production (PROD), pharmaceutical employment (EMP), pharmaceutical market value (MV), pharmaceutical export (EXP), and pharmaceutical import (IMP) were the independent variables. The relationship between innovation performance and life expectancy was analysed using Spearman's correlation.

## 4. Results and discussion

First, the data were cleaned to test the linear regression conditions and the descriptive statistics of the variables examined to detect erroneous values (Table 1). Missing and/or erroneous values were not included in the raw dataset.

**Table 1** Descriptive statistics

|  | GII | R&D | Production | Employment | Market Value | Export | Import | Life Expectancy |
|---|---|---|---|---|---|---|---|---|
| Valid | 27 | 27 | 27 | 27 | 27 | 27 | 27 | 27 |
| Missing | 0 | 0 | 0 | 0 | 0 | 0 | 0 | 0 |
| Mean | 47.663 | 1399.704 | 10847.000 | 32779.259 | 8326.481 | 15662.741 | 11980.296 | 80.488 |
| Std. Deviation | 9.009 | 2259.674 | 14153.496 | 29849.172 | 10170.120 | 23247.149 | 14194.740 | 2.953 |
| Shapiro-Wilk | 0.953 | 0.654 | 0.760 | 0.858 | 0.719 | 0.699 | 0.769 | 0.887 |
| P-value of Shapiro-Wilk | 0.252 | <.001 | <.001 | 0.002 | <.001 | <.001 | <.001 | 0.007 |
| Minimum | 34.900 | 38.000 | 121.000 | 1755.000 | 177.000 | 333.000 | 312.000 | 73.200 |
| Maximum | 66.100 | 8466.000 | 54305.000 | 119994.000 | 40456.000 | 81862.000 | 52679.000 | 84.000 |

The independent variables were not normally distributed as the P-values in the Shapiro-Wilk test as shown in Table 1; thus, the log transformation was used to transform skewed data to close to 'normal' and augment the reliability of the analyses. Table 2 shows the descriptive statistics of the new independent variables, which are now normally distributed.

**Table 2** Descriptive Statistics

|  | logR&D | logPROD | logEMP | logMV | logEXP | logIMP |
|---|---|---|---|---|---|---|
| Valid | 27 | 27 | 27 | 27 | 27 | 27 |
| Missing | 0 | 0 | 0 | 0 | 0 | 0 |
| Mean | 2.594 | 3.580 | 4.314 | 3.638 | 3.648 | 3.754 |
| Std. Deviation | 0.7273 | 0.7268 | 0.4762 | 0.5343 | 0.7639 | 0.5799 |
| Shapiro-Wilk | 0.9346 | 0.9664 | 0.9629 | 0.9640 | 0.9335 | 0.9635 |
| P-value of Shapiro-Wilk | 0.090 | 0.510 | 0.428 | 0.454 | 0.084 | 0.441 |
| Minimum | 1.580 | 2.083 | 3.244 | 2.248 | 2.522 | 2.494 |
| Maximum | 3.928 | 4.735 | 5.079 | 4.607 | 4.913 | 4.722 |

The boxplots showed that there were no outliers in the database. Due to the strong correlation with pharmaceutical R&D, pharmaceutical production was removed from the model. Pharmaceutical MV was also not included in the final model because it is highly correlated with pharmaceutical employment (Table 3). The third variable not included is pharmaceutical import because it is highly correlated with pharmaceutical export. As a result, only three independent variables remained in the final model (pharmaceutical R&D, pharmaceutical EMP, and pharmaceutical EXP).

**Table 3** Pearson's Correlations

| Variable | | logR&D | logPROD | logEMP | logMV | logEXP | logIMP |
|---|---|---|---|---|---|---|---|
| 1. logR&D | Pearson's r | — | | | | | |
| 2. logPROD | Pearson's r | **0.899 \*\*\*** | — | | | | |
| 3. logEMP | Pearson's r | 0.707 \*\*\* | 0.752 \*\*\* | — | | | |
| 4. logMV | Pearson's r | 0.709 \*\*\* | 0.746 \*\*\* | **0.837 \*\*\*** | — | | |
| 5. logEXP | Pearson's r | 0.740 \*\*\* | 0.723 \*\*\* | 0.541 \*\* | 0.428 \* | — | |
| 6. logIMP | Pearson's r | 0.694 \*\*\* | 0.701 \*\*\* | 0.703 \*\*\* | 0.621 \*\*\* | **0.850 \*\*\*** | — |

\* $p < .05$, \*\* $p < .01$, \*\*\* $p < .001$

The scatter plot of residuals vs. predicted and the standardised residual histogram (Figure 9) imply that the linear relationship and homoscedasticity conditions are satisfied.

**Figure 9** Scatter plot residuals vs predicted, and the standardised residual histogram

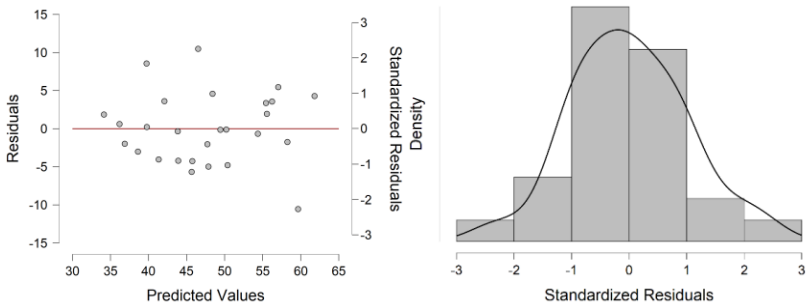

The Durbin-Watson d value of 1.761, which is close to the ideal value of 2.0, signals no disturbing autocorrelation in the sample. Figure 10 shows the positions of the countries, determined by their scores on the global innovation index (GII) and their logarithmical scores of pharmaceutical R&D, pharmaceutical exports, and pharmaceutical employment data.

**Figure 10** Scree-plot of the countries (Global Innovation Index and pharmaceutical R&D, pharmaceutical export and pharmaceutical employment)

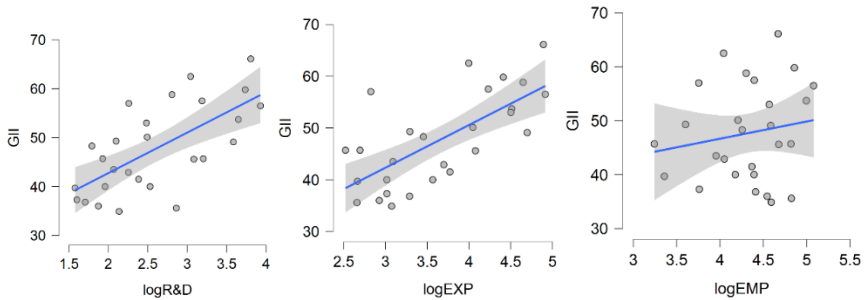

The ANOVA table (Table 4) shows that the model is significant, and therefore generalisable to the population.

**Table 4** ANOVA table

| Model | | Sum of Squares | df | Mean Square | F | p |
|---|---|---|---|---|---|---|
| $H_1$ | Regression | 1549.021 | 3 | 516.340 | 21.166 | <.001 |
| | Residual | 561.082 | 23 | 24.395 | | |
| | Total | 2110.103 | 26 | | | |

*Note.* The intercept model is omitted since no meaningful information can be shown.

As shown in Table 5, $R^2$ value of 0.734 indicates that the resulting model is robust, with 73.4% of the variance in the output variable (GII) explained by the predictor variables (logR&D, logEXP, and logEMP). The adjusted $R^2$ indicates that the explanatory power of the model for the population is also very high (69.9%). It suggests that only three pharmaceutical indicators could explain the innovation performance of a country.

This supports the findings of Duenas-Gonzalez and Gonzalez-Fierro (2020) that pharmaceutical innovation leads to economic growth and is a sign of a country's innovation efforts.

**Table 5** Model Summary - GII

| Model | R | R² | Adjusted R² | RMSE | Autocorrelation | Durbin-Watson Statistic | p |
|---|---|---|---|---|---|---|---|
| H₀ | 0.000 | 0.000 | 0.000 | 9.009 | 0.203 | 1.522 | 0.202 |
| H₁ | 0.857 | 0.734 | 0.699 | 4.939 | 0.108 | 1.761 | 0.560 |

Table 6 shows that the global innovation performance of a country can be estimated using the following formula:

$$GII = 53.805 + 9.389 * logR\&D - 11.819 * logEMP + 5.616 * logEXP$$

Standardised beta coefficients suggest that pharmaceutical R&D, pharmaceutical employment, and pharmaceutical exports strongly influence innovation performance; therefore, the research hypothesis is accepted.

**Table 6** Coefficients

| Model | | Unstandardised | Standard Error | Standardised | t | p |
|---|---|---|---|---|---|---|
| H₀ | (Intercept) | 47.663 | 1.734 | | 27.491 | <.001 |
| H₁ | (Intercept) | 53.805 | 10.031 | | 5.364 | <.001 |
| | logR&D | 9.389 | 2.359 | 0.758 | 3.980 | <.001 |
| | logEMP | -11.819 | 2.880 | -0.625 | -4.104 | <.001 |
| | logEXP | 5.616 | 1.887 | 0.476 | 2.976 | 0.007 |

Pharmaceutical R&D and pharmaceutical exports have a positive effect on innovation performance, while higher employment in the pharmaceutical industry has a negative effect on it. This means that in countries where pharmaceutical R&D spending is high and pharmaceutical export is significant, we can generally expect high innovation performance, especially when pharmaceutical labour intensity is low. In other words, per capita value-added in the pharmaceutical sector should be high for a country to perform better in global innovation competition. Highly skilled labour and automated R&D activities supported by artificial intelligence in the pharmaceutical sector have a positive impact on a country's global innovation performance.

The relationship between innovation performance and life expectancy is also examined. Since the life expectancy variable has a non-normal distribution, Spearman correlation was used to indicate the direction of the relationship between the two variables. The Spearman correlation coefficient (rs=0.661) indicates a strong positive relationship between the global innovation index and life expectancy (Table 7). Therefore, we can assume that when a country's innovation performance increases, life expectancy also increases; that is, innovation positively influences life expectancy. Understanding this relationship is particularly important for Hungary, where life expectancy is among the lowest in OECD countries (Uzzoli, 2016).

As shown by both the MLR and Spearman's correlation results, the development of the pharmaceutical industry in a country has a positive impact on innovation performance and life expectancy. Our findings support previous research that innovation (Khullar, Fisher, and Chandra 2019), particularly pharmaceutical innovation (Omachonu and Einspruch 2010, Lichtenberg 2012), has a positive impact on life expectancy. Pharmaceutical innovation is particularly important because Lichtenberg (2012), in a study that examined 30 developing and developed countries between 2000 and 2009, found that it was responsible for 73 percent of the increase in life expectancy. Pharmaceutical innovation, the introduction and use of cost-effective new medicines, significantly increased cancer survival rates in New Zealand and significantly reduced premature cancer mortality in the period to 1998-2017 (Lichtenberg 2021).

**Table 7** Spearman's Correlations

| Variable |  | Life Expectancy | GII |
|---|---|---|---|
| 1. Life Expectancy | Spearman's rho | — | |
| 2. GII | Spearman's rho | 0.661 *** | — |

\* p <.05, \*\* p <.01, \*\*\* p <.001

## 5. Conclusions

Due to the pandemic, the role of the pharmaceutical industry has never been more important. More specifically, new treatments have not been developed in such a short time, as in the last two years. This suggests that the pharmaceutical industry needs to become increasingly innovative, with practical examples to show, such as the development of genetically engineered state-of-the-art RNA and DNA vaccines for COVID-19. Therefore, innovation and R&D are becoming increasingly critical in this sector. At the same time, our results indicate that an innovative pharmaceutical industry has a strong positive impact on a country's overall innovation performance. It creates a knowledge-based ecosystem that has a positive impact on innovation in other industries. Based on our research findings, we recommend that economic policymakers who want to improve a country's innovation performance use all possible means to promote the establishment and development of an innovative pharmaceutical industry. In particular, the promotion of pharmaceutical R&D and exports play an important role in the development of an innovation ecosystem. Our results also show that the number of employees in the pharmaceutical industry negatively impacts innovation performance. Therefore, we can assume that innovation performance improves in countries where a smaller number of employees create more added value in the pharmaceutical industry. In this respect, Europe, in particular, lags far behind North America, where the indicator of gross value-added per employee in the pharmaceutical industry is 2.16 times higher (Statista 2012). This indicates a more distant prospect and calls for conscious development of knowledge using the latest technology (e.g. with the help of artificial intelligence). Finally, the development of an innovative pharmaceutical industry has a positive impact on life expectancy. Since our research results focused on the European context, our future research direction is to extend this research to other countries around the world.

# INTERNET SOURCES

## Appendix 1 – Global Innovation Index ranking

| | | | | | | | | | | |
|---|---|---|---|---|---|---|---|---|---|---|
| 1 | Switzerland | 66.1 | | 61 | Armenia | 32.6 | | 121 | Algeria | 19.5 |
| 2 | Sweden | 62.5 | | 62 | Brazil | 31.9 | | 122 | Zambia | 19.4 |
| 3 | United States of America | 60.6 | | 63 | Georgia | 31.8 | | 123 | Mali | 19.2 |
| 4 | United Kingdom | 59.8 | | 64 | Belarus | 31.3 | | 124 | Mozambique | 18.7 |
| 5 | Netherlands | 58.8 | | 65 | Tunisia | 31.2 | | 125 | Togo | 18.5 |
| 6 | Denmark | 57.5 | | 66 | Saudi Arabia | 30.9 | | 126 | Benin | 18.1 |
| 7 | Finland | 57 | | 67 | Iran | 30.9 | | 127 | Ethiopia | 18.1 |
| 8 | Singapore | 56.6 | | 68 | Colombia | 30.8 | | 128 | Niger | 17.8 |

| | | | | | | | | | |
|---|---|---|---|---|---|---|---|---|---|
| 9 | Germany | 56.5 | 69 | Uruguay | 30.8 | 129 | Myanmar | 17.7 |
| 10 | Republic of Korea | 56.1 | 70 | Qatar | 30.8 | 130 | Guinea | 17.3 |
| 11 | Hong Kong. China | 54.2 | 71 | Brunei | 29.8 | 131 | Yemen | 13.6 |
| 12 | France | 53.7 | 72 | Jamaica | 29.1 | | | |
| 13 | Israel | 53.5 | 73 | Panama | 29 | | | |
| 14 | China | 53.3 | 74 | Bosnia and Herzegovina | 29 | | | |
| 15 | Ireland | 53 | 75 | Morocco | 29 | | | |
| 16 | Japan | 52.7 | 76 | Peru | 28.8 | | | |
| 17 | Canada | 52.3 | 77 | Kazakhstan | 28.6 | | | |
| 18 | Luxembourg | 50.8 | 78 | Kuwait | 28.4 | | | |
| 19 | Austria | 50.1 | 79 | Bahrain | 28.4 | | | |
| 20 | Norway | 49.3 | 80 | Argentina | 28.3 | | | |
| 21 | Iceland | 49.2 | 81 | Jordan | 27.8 | | | |
| 22 | Belgium | 49.1 | 82 | Azerbaijan | 27.2 | | | |
| 23 | Australia | 48.4 | 83 | Albania | 27.1 | | | |
| 24 | Czech Republic | 48.3 | 84 | Oman | 26.5 | | | |
| 25 | Estonia | 48.3 | 85 | Indonesia | 26.5 | | | |
| 26 | New Zealand | 47 | 86 | Kenya | 26.1 | | | |
| 27 | Malta | 46.4 | 87 | Lebanon | 26 | | | |
| 28 | Italy | 45.7 | 88 | Tanzania | 25.6 | | | |
| 29 | Cyprus | 45.7 | 89 | Botswana | 25.4 | | | |
| 30 | Spain | 45.6 | 90 | Dominican Rep. | 25.1 | | | |
| 31 | Portugal | 43.5 | 91 | Rwanda | 25.1 | | | |
| 32 | Slovenia | 42.9 | 92 | El Salvador | 24.8 | | | |
| 33 | Malaysia | 42.4 | 93 | Uzbekistan | 24.5 | | | |
| 34 | United Arab Emirates | 41.8 | 94 | Kyrgyzstan | 24.5 | | | |
| 35 | Hungary | 41.5 | 95 | Nepal | 24.4 | | | |
| 36 | Latvia | 41.1 | 96 | Egypt | 24.2 | | | |
| 37 | Bulgaria | 40 | 97 | Paraguay | 24.1 | | | |
| 38 | Poland | 40 | 98 | Trinidad and Tobago | 24.1 | | | |
| 39 | Slovakia | 39.7 | 99 | Ecuador | 24.1 | | | |
| 40 | Lithuania | 39.2 | 100 | Cabo Verde | 23.9 | | | |
| 41 | Croatia | 37.3 | 101 | Sri Lanka | 23.8 | | | |
| 42 | Viet Nam | 37.1 | 102 | Senegal | 23.7 | | | |
| 43 | Greece | 36.8 | 103 | Honduras | 23 | | | |

| 44 | Thailand | 36.7 | 104 | Namibia | 22.5 | | | |
|---|---|---|---|---|---|---|---|---|
| 45 | Ukraine | 36.3 | 105 | Bolivia | 22.4 | | | |
| 46 | Romania | 36 | 106 | Guatemala | 22.4 | | | |
| 47 | Russian Federation | 35.6 | 107 | Pakistan | 22.3 | | | |
| 48 | India | 35.6 | 108 | Ghana | 22.3 | | | |
| 49 | Montenegro | 35.4 | 109 | Tajikistan | 22.2 | | | |
| 50 | Philippines | 35.2 | 110 | Cambodia | 21.5 | | | |
| 51 | Turkey | 34.9 | 111 | Malawi | 21.4 | | | |
| 52 | Mauritius | 34.4 | 112 | Côte d'Ivoire | 21.2 | | | |
| 53 | Serbia | 34.3 | 113 | Lao PDR | 20.6 | | | |
| 54 | Chile | 33.9 | 114 | Uganda | 20.5 | | | |
| 55 | Mexico | 33.6 | 115 | Madagascar | 20.4 | | | |
| 56 | Costa Rica | 33.5 | 116 | Bangladesh | 20.4 | | | |
| 57 | North Macedonia | 33.4 | 117 | Nigeria | 20.1 | | | |
| 58 | Mongolia | 33.4 | 118 | Burkina Faso | 20 | | | |
| 59 | Republic of Moldova | 33 | 119 | Cameroon | 20 | | | |
| 60 | South Africa | 32.7 | 120 | Zimbabwe | 20 | | | |

## Appendix 2 – Raw dataset for analysis

| Country | GII | R&D | Production | Employment | Market Value | Export | Import | Life Exp. |
|---|---|---|---|---|---|---|---|---|
| | 0-100 | EUR million | EUR million | units | EUR million | EUR million | EUR million | years |
| Austria | 50.1 | 311 | 3024 | 16094 | 4583 | 11150 | 9898 | 82 |
| Belgium | 49.1 | 3846 | 17547 | 38489 | 5988 | 49732 | 42332 | 82.1 |
| Bulgaria | 40 | 91 | 121 | 15000 | 1210 | 1038 | 1471 | 74.96 |
| Croatia | 37.3 | 40 | 664 | 5763 | 957 | 1039 | 1326 | 78.07 |
| Cyprus | 45.7 | 85 | 253 | 1755 | 177 | 333 | 312 | 80.38 |
| Czech Rep | 48.3 | 62 | 858 | 18000 | 3010 | 2852 | 4960 | 79.3 |
| Denmark | 57.5 | 1543 | 14391 | 24821 | 3111 | 17041 | 4217 | 81.5 |

| Finland | 57 | 182 | 1877 | 5672 | 2712 | 669 | 1985 | 82.1 |
| France | 53.7 | 4451 | 35848 | 98780 | 29304 | 32556 | 26012 | 82.9 |
| Germany | 56.5 | 8466 | 33158 | 119994 | 40456 | 81862 | 52679 | 81.4 |
| Greece | 36.8 | 51 | 1376 | 25700 | 5158 | 1944 | 2957 | 81.7 |
| Hungary | 41.5 | 242 | 3859 | 23300 | 2631 | 5916 | 4688 | 76.4 |
| Ireland | 53 | 305 | 19305 | 37000 | 2279 | 31666 | 27867 | 82.8 |
| Italy | 45.7 | 1600 | 34000 | 65800 | 24099 | 498 | 665 | 83.6 |
| Netherlands | 58.8 | 642 | 6180 | 20000 | 5770 | 44382 | 29928 | 82.2 |
| Norway | 49.3 | 126 | 1072 | 4000 | 2621 | 1964 | 1382 | 83 |
| Poland | 40 | 339 | 2550 | 24736 | 7281 | 3688 | 6906 | 78 |
| Portugal | 43.5 | 117 | 1737 | 9000 | 3409 | 1229 | 2803 | 81.8 |
| Romania | 36 | 75 | 655 | 35000 | 3130 | 845 | 3473 | 75.36 |
| Russia | 35.6 | 727 | 5881 | 66523 | 17678 | 457 | 13579 | 73.2 |
| Slovakia | 39.7 | 38 | 356 | 2287 | 1455 | 462 | 1865 | 77.8 |
| Slovenia | 42.9 | 180 | 1659 | 11213 | 675 | 4985 | 4002 | 81.6 |
| Spain | 45.6 | 1212 | 15832 | 47449 | 17105 | 11953 | 14767 | 83.9 |
| Sweden | 62.5 | 1104 | 9840 | 11012 | 4313 | 9918 | 4391 | 83.2 |
| Switzerland | 66.1 | 6383 | 54305 | 46652 | 5533 | 77809 | 29036 | 84 |
| Turkey | 34.9 | 137 | 3482 | 39000 | 6891 | 1189 | 4598 | 78.6 |
| United Kingdom | 59.8 | 5437 | 23039 | 72000 | 23279 | 25717 | 25369 | 81.3 |

Source: based on Cornell University, INSEAD, and WIPO (2020); EFPIA (2021); STATISTA (2017)